\documentclass[twocolumn,showpacs,showkeys,preprintnumbers,amsmath,amssymb,prb]
{revtex4}
\usepackage{graphicx}
\usepackage{dcolumn}
\usepackage{bm}

\begin{document}

\title{Thermodynamic properties of a tetramer Ising-Heisenberg bond alternating chain \\
as a model system for Cu(3-Chloropyridine)$_2$(N$_3$)$_2$}
\author{Jozef Stre\v{c}ka}
\email{jozkos@pobox.sk}
\author{Michal Ja\v{s}\v{c}ur}
\affiliation{Department of Theoretical Physics and Astrophysics, 
Faculty of Science, \\ P. J. \v{S}af\'{a}rik University, Park Angelinum 9,
040 01 Ko\v{s}ice, Slovak Republic}
\author{Masayuki Hagiwara}
\affiliation{KYOKUGEN (Research Center for Materials Science at Extreme Conditions), 
Osaka University, 1-3 Machikaneyama, Toyonaka, Osaka 560-8531, Japan}
\author{Kazuhiko Minami}
\affiliation{Graduate School of Mathematics, Nagoya University,                  
             Nagoya 464-8602, Japan}
\author{Yasuo Narumi}
\author{Koichi Kindo}
\affiliation{Institute for Solid State Physics, University of Tokyo, 
             Kashiwa, Chiba 277-8581, Japan}

\date{\today}

\begin{abstract}
Thermodynamic properties of a tetramer ferro-ferro-antiferro-antiferromagnetic
Ising-Heisen\-berg bond alternating chain are investigated by the use of an 
exact mapping transformation technique. Exact results for the magnetization, susceptibility and specific heat in the zero as well as nonzero magnetic 
field are presented and discussed in detail. 
The results obtained from the mapping are compared with the relevant 
experimental data of Cu(3-Clpy)$_2$(N$_3$)$_2$ (3-Clpy=3-Chloropyridine).
\end{abstract}


\maketitle

\section{Introduction}

Quantum behaviour of low-dimensional molecular-based magnetic materials 
has become one of the most fascinating topics emerging at the border 
of condensed matter physics, materials science, and inorganic chemis\-try. 
In this area, the quantum ferrimagnetic chains (QFC) have attracted 
considerable attention during the last few years, because they exhibit a remarkable 
combination of ferro\-magnetic (F) and antiferromagnetic (AF) features. \cite{Pati,Yama} 
In an attempt directed toward the synthesis and design 
of possible experimental realizations of QFC, a magnetostructural 
analysis of several bimetallic assemblies has been accomplished, \cite{1Dbime} 
since the mixed-spin chains afford their most simple and common representatives. 
Up to now, the dual aspect of QFC has been experimentally  
confirmed in the [NiCu(pba)(D$_2$O)$_3$].2D$_2$O \cite{QFC1} and 
[MnCu(mal)$_2$(H$_2$O)$_4$].2H$_2$O \cite{QFC2} mixed-spin chains.  

In addition to the mixed-spin chains, another class of the QFC constitute 
the bond alternating chains (BAC) with an unusual fashion of exchange bonds. 
From the experimental point of view, the bond alternation in the 1D polymeric assemblies 
demands at least two structurally nonequivalent superexchange 
paths in order to get a series of alternating exchange bonds. It should be mentioned, however, that bond alternation may also arise in a system with an uniform superexchange pathway as a result of the spin-Peierls phenomenon, \cite{CF} spontaneous dimerization, which occurs when the elastic energy increase connected with the lattice distortion is lower than the corresponding magnetic energy gain arising from the dimerization. Hence, the recent discovery of the first inorganic spin-Peierls compounds CuGeO$_3$ \cite{CUGEO} and $\alpha^{'}$-NaV$_2$O$_5$ \cite{NAVO} gave rise to a number of theoretical works devoted to the spin-1/2 AF-AF BAC. \cite{SP} 

Another exciting field in molecular magnetism was opened up by Haldane's conjecture, \cite{Haldane} which has already been experimentally verified in several AF nickel-based chains (for a review of Haldane gap compounds see Ref. \onlinecite{HalExp}).
As firstly pointed out by Hida, \cite{FAFtheor} additional insight into the striking properties of the spin-1 AF chain can be acquired by analyzing the spin-1/2 F-AF BAC in the strong-F-coupling limit. \cite{FAFtheo} Accordingly, much effort has been addressed to prepare polymeric complexes, in which the F-AF bond alternation should be realized. At present, there exist several copper-based polymeric compounds that fulfill this requirement \cite{FAFexp, FAFexp0, FAFexp1, FAFexp2, FAFexp3, FAFexp4, FAFexp5} (see Table I) and  Haldane-like behavior has indeed been undoubtedly proved to occur in (IPA)CuCl$_3$. \cite{FAFexp1} 

\begin{table}
\caption{Several examples of one-dimensional copper-based chains with alternating 
         F and AF interactions.}
\begin{ruledtabular}         
\begin{tabular}{lcr} 
Chemical formula & Bond alternation & Ref. \\ \hline
$[$Cu(TIM)$]$CuCl$_4$ & F-AF & [\onlinecite{FAFexp}] \\ 
(4-BzPip)CuCl$_3$ & F-AF & [\onlinecite{FAFexp0}] \\ 
(IPA)CuCl$_3$ & F-AF & [\onlinecite{FAFexp1}] \\ 
(DMA)CuCl$_3$ & F-AF & [\onlinecite{FAFexp2}] \\  
$[$Cu(bipy)(N$_3$)$_2]$ & F-AF & [\onlinecite{FAFexp3}] \\ 
$[$Cu$_2$(Me$_2$Eten)$_2$(N$_3$)$_2]$ (N$_3$)$_2$ & F-AF & [\onlinecite{FAFexp4}] \\ 
$[$Cu$_2$(ampy)$_2$(N$_3$)$_2]$ (N$_3$)$_2$ & F-AF & [\onlinecite{FAFexp5}] \\ \hline 3CuCl$_2$.2Dx & F-F-AF & [\onlinecite{FFAFexp}] \\ \hline 
Cu(3-Clpy)$_2$(N$_3$)$_2$ & F-F-AF-AF & [\onlinecite{FFAFAF}] \\ 
\end{tabular} 
\end{ruledtabular}
\footnotetext{\textit{Abbreviations}: TIM = 2,3,9,10-tetramethyl-1,3,8,10-tetraene\-cyclo-1,4,8,11-tetraazatetradecane;
4-BzPip = 4-benzyl\-piperi\-dinium (1+) ion; IPA = isopropylammonium (1+) ion; 
DMA = dimethyl\-ammonium (1+) ion; bipy = bipyridine; Dx = 1,4-dioxane; Me$_2$Eten = N,N-dimethyl-N$^{'}$-ethylethylendiamine; ampy = 1-(2-aminoethyl)pyrrolidine.}
\end{table}

Of particular interest are also other BAC with more peculiar bond alternation, especially with a longer repeating unit of exchange bonds; namely, according to the Oshikawa-Yamanaka-Affleck rule, \cite{OYA} one may expect the appearance of the magnetization plateau in any system with a longer period of the ground state. 
Of course, this rule represents just a necessary condition for the plateau-state formation and does not directly prove its existence in any specific model. The theoretical investigations focused on the magnetization process of the spin-1/2 F-F-AF BAC thus revealed another interesting finding - the breakdown of the magnetization plateau. \cite{FFAFtheor} When the ratio between F  and AF coupling constants is strong enough in this system, the plateau state disappears from the magnetization curve. In agreement with this finding, there has not been found any plateau in the low-temperature magnetization curve of 3CuCl$_2$.2Dx, \cite{FFAFexp} which is regarded as a typical example of the spin-1/2 F-F-AF BAC with the strong F and weak AF coupling. 

The versatility of the azido ligand in bridging the magnetic ions in various fashions is
nicely demonstrated in the copper-based compound Cu(3-Clpy)$_2$(N$_3$)$_2$, \cite{FFAFAF} hereafter abbreviated as CCPA. There is a strong evidence that this compound can be regarded as a spin-1/2 tetramer chain with the F-F-AF-AF bond alternation. \cite{Hagiwara1, Hagiwara2, Hagiwara3, Hagiwara4} 
However, the 1D nature of the CCPA can be attributed to a sufficient separation between chains, which is ensured by the large steric hindrance of the bulky 3-Clpy ligands (see Fig. 1a). The peculiar F-F-AF-AF sequence of the exchange bonds arises, on the other hand, on account of two kinds of bridging fashions of the azido group: the magnetic Cu$^{2+}$ ions are linked either in a double end-on, or in an end-on and end-to-end bridging fashion. It is quite well established \cite{Az} that the end-on bridges are usually associated with the F coupling (with exception when the bridging angle is too large), while the end-to-end bridges are responsible for the AF coupling. The complete magnetic studies of the CCPA have been performed by Hagiwara and co-workers: for a powder sample high-field magnetization and the susceptibility measurements were reported, \cite{Hagiwara1} while for a single-crystal sample the high-field magnetization and, susceptibility are known, \cite{Hagiwara2} as well as the specific heat \cite{Hagiwara3} and electron-spin resonance \cite{Hagiwara4} data.

The primary purpose of this work is to provide a detailed description to the thermodynamic properties of the spin-1/2 F-F-AF-AF BAC by means of a simplified  Ising-Heisenberg model suggested in our preliminary report.\cite{ccpa} Due to the simplicity of the proposed Hamiltonian as well as its low-dimensionality, an accurate analytical treatment for the complete set of thermodynamic quantities can be elaborated within an exact mapping transformation technique.\cite{DIT} To the best of our know\-ledge, there have been reported only a few rigorous results for the spin-1/2 F-F-AF-AF BAC obtained by applying the exact diagonalization method for a finite-size Heisenberg cluster \cite{Hagiwara1, Hagiwara2, Hagiwara3, Hagiwara4} and any further more comprehensive studies have not been reported in the literature hitherto.\cite{Lu}

The outline of this paper is as follows. In the next section, we shall provide the detailed description of the model system and then, the basic ideas of transformation procedure will be presented. This is followed by the presentation of the most interesting results. An exhaustive survey of results for several thermodynamic quantities in the zero as well as non-zero external field is reviewed in Section IIIA, 
while a comparison with the relevant experimental data is included in Section IIIB. Finally, some concluding remarks are drawn in Section IV.

\section{Model and method}

A fragment of the CCPA crystal structure is depicted in Fig. 1a. Apparently, there are three nonequivalent positions of the Cu$^{2+}$ ions due to two different kinds of azido bridges. To match the structural situation from the magnetic viewpoint, Fig. 1b schematically reproduces the magnetic structure of CCPA: 
the sites interacting purely via the F interaction $J_F$ are denoted as the Cu1 sites, the sites coupled via both F as well as AF interaction as the Cu2 sites, and finally, the sites coupled purely through the AF interaction $J_{AF}$ are labeled as the Cu3 sites. As a consequence of the structural differences, one 
also has to assume various $g$-factors $g_1$, $g_2$, and $g_3$ at the 
Cu1, Cu2, and Cu3 sites, respectively.

It is worthwhile to say that the supposed magnetic structure can also be identified as a ferromagnetic chain, the bonds of which are decorated by the AF trimers 
(Fig. 1b). Owing to this fact, it is very advisable to assume that 
the F interaction $J_F$ has an Ising-type character, whereas the AF interaction $J_{AF}$ may have the more general form of an anisotropic Heisenberg coupling (the detailed discussion will be given in Section III). Under the circumstances, the model under consideration can be exactly treated by applying a generalized decoration-iteration mapping transformation. \cite {ccpa} Actually, this mapping procedure has been proved to be very useful in investiga\-ting several mixed-bond Ising-Heisenberg models, some of the present authors already obtained within this scheme exact results for the trimerized Ising-Heisenberg linear chain, \cite{LC} the Ising-Heisenberg diamond chain \cite{DC}, and some decorated 
Ising-Heisenberg planar models. \cite{2D} In what follows, we will refer to the 
Cu1 magnetic sites as to the Ising-type sites, whereas the Cu2 and Cu3 sites will
be denoted as the Heisenberg-type sites.

Let us write the total Hamiltonian of the spin-1/2 F-F-AF-AF Ising-Heisenberg
chain comprised of $4N$ magnetic sites. By imposing a periodic boundary condition ($S_{4N + 1} = S_{1}$), 
the total Hamiltonian of the system takes the form:	 
\begin{eqnarray}
	   H \! \! &=& \! \! J_{AF} \sum_{k=1}^{N} 
	[(\textbf{S}_{4k-3} , \textbf{S}_{4k-2})_{\Delta} 
	 + (\textbf{S}_{4k-2} , \textbf{S}_{4k-1})_{\Delta}] \nonumber \\
	\! \! &-& \! \!  J_{F} \sum_{k=1}^{N} 
	  [S_{4k-1}^{z} S_{4k}^{z} + S_{4k}^{z} S_{4k+1}^{z}]
   - B_1 \sum_{k = 1}^{N} S_{4k}^{z}          \nonumber \\
  \! \! &-& \! \! B_2 \sum_{k = 1}^{N} (S_{4k-1}^{z} + S_{4k-3}^{z})
   - B_3 \sum_{k = 1}^{N} S_{4k-2}^{z}, 
\label{r1}	   
\end{eqnarray}
where $(\textbf{S}_i, \textbf{S}_j)_{\Delta} = \Delta (S_i^x S_j^x + S_i^y S_j^y) 
+ S_i^z S_j^z$, $S_i^{\alpha}$ $(\alpha = x,y,z)$ marks the spatial components of the spin-1/2 operator at the $i$th lattice point and various on-site magnetic fields $B_j = g_j \mu_B B$ $(j=1,2,3)$ have been introduced in order to distinguish 
the $g$-factors at the Cu1, Cu2 and Cu3 sites. The first summation accounts for the  nearest-neighbor AF Heisenberg coupling ($J_{AF} > 0$), $\Delta$ is the spatial anisotropy in this interaction, and the second summation accounts the nearest-neighbor F Ising coupling ($J_{F} > 0$). Other quantities have the usual meaning: $\mu_B$ is the Bohr magneton and $B$ the external magnetic field.

For convenience, the total Hamiltonian (\ref{r1}) can be rewritten as a sum of bond Hamiltonians $H = \sum_{k} H_k$, where  each bond Hamiltonian $H_k$ involves all the interaction terms associated with the $k$th AF Heisenberg trimer (Fig. 1b):
\begin{eqnarray}
	 \textsl{H}_k \! \! &=& \! \! J_{AF} 
	  [(\textbf{S}_{4k-3} , \textbf{S}_{4k-2})_{\Delta} 
	 + (\textbf{S}_{4k-2} , \textbf{S}_{4k-1})_{\Delta}] \nonumber \\
\! \! &-& \! \! J_{F} (S_{4k-4}^{z} S_{4k-3}^{z} + S_{4k-1}^{z} S_{4k}^{z}) - B_3  S_{4k-2}^{z}  \nonumber \\
\! \! &-& \! \! B_2 (S_{4k-1}^{z} + S_{4k-3}^{z}) 
       - B_1 (S_{4k-4}^{z} + S_{4k}^{z})/2.
\label{r2}	   
\end{eqnarray}
Obviously, the different bond Hamiltonians commute with respect to each other, hence, the partition function $Z$ can be partially factorized into products of the bond partition functions $Z_{k}$:
\begin{eqnarray}
Z \! \! &=& \! \! \mbox{Tr}_{ \{ 4, 8, ...4N \}} 
\prod_{k=1}^{N} \mbox{Tr}_{ \{ 4k-3, 4k-2, 4k-1 \}} \exp(- \beta H_k) \nonumber \\
Z \! \! &=& \! \! \mbox{Tr}_{ \{ 4, 8, ...4N \}} \prod_{k=1}^{N} Z_k. 
\label{r3}	
\end{eqnarray}
In the above, $\beta = 1/(k_B T)$, $k_B$ is Boltzmann's constant, $T$ is the absolute temperature, the symbol Tr$_{\{ 4k-3, 4k-2, 4k-1 \}}$ means the trace over the $k$th AF Heisenberg trimer, and Tr$_{ \{ 4, 8, ..., 4N \}}$ stands for the trace over the spin states of all Ising-type (Cu1) spins. To proceed further with the calculation, one may introduce at the level of the bond partition function $Z_k$ an extended decoration-iteration transformation by adopting the same idea as already discussed in several papers based on this mapping scheme (see for instance Refs. [\onlinecite{LC}]-[\onlinecite{2D}]):
\begin{widetext}
\begin{eqnarray}
&& \! \! \! Z_k = \mbox{Tr}_{ \{ 4k-3, 4k-2, 4k-1 \}} \exp(- \beta H_k) =      
   \exp \Bigl[\beta B_1 (S_{4k-4}^{z} + S_{4k}^{z})/2 \Bigr] 
   \biggl \{ \exp[\beta (J_{AF} + J_{1}^{+})/6] \sum_{n=1}^{3} \exp(\beta x_n)+
 \label{r4}  \\
&& \! \! \! \exp[\beta (J_{AF} - J_{1}^{+})/6] \sum_{n=1}^{3} \exp(\beta y_n) 
   + 2 \exp(- \beta J_{AF}/2) \cosh (\beta J_{1}^{+}/2)\biggr \} 
   = A \exp[ \beta R S_{4k-4}^{z} S_{4k}^{z} + \beta H_0 (S_{4k-4}^{z} +   
             S_{4k}^{z})/2]. \nonumber
\end{eqnarray}
\end{widetext}
Since a complete explicit rendering of the decoration-iteration mapping is in this case rather intricate, we have defined in the transformation Eq. (\ref{r4}) several functions in order to write it in a more appropriate form:
\begin{eqnarray}
J_{1}^{+} \! \! &=& \! \! J_F (S_{4k-4}^{z} + S_{4k}^{z})+ 2 B_2 + B_3,   
\label{e4}   \\
J_{2}^{+} \! \! &=& \! \! J_F (S_{4k-4}^{z} + S_{4k}^{z})+ 2 (B_2 - B_3), 
\label{ee4}  \\
J_{1}^{-} \! \! &=& \! \! J_F (S_{4k-4}^{z} - S_{4k}^{z}),                
\label{eee4}	
\end{eqnarray}
and the terms $x_n$ and $y_n$ denote the roots of two cubic equations that 
are given as follows:
\begin{eqnarray}
\! \! \! x_n \! \! &=& \! \! \pm 2 \sqrt{P_1} \cos[\Phi_1 + (n-1)2 \pi/3], 
                              \> (n=1,2,3), \\
\! \! \! P_1 \! \! &=& \! \! [(J_{AF} + J_2^{+})/6]^2 + 
                             [(J_1^{-})^2 + 2 (J_{AF} \Delta)^2]/12,  \\
\! \! \! Q_1 \! \! &=& \! \! [(J_{AF} + J_2^{+})/6]^3 +  \nonumber \\
      &&  \quad  +   [(J_{AF} + J_2^{+})/6] [(J_{AF} \Delta)^2 - (J_1^{-})^2]/4, \\
\! \! \! \Phi_1 \! \! &=& \! \! 
         \frac13 \arctan \Bigl( \sqrt{P_1^3 - Q_1^2}/Q_1 \Bigr); 
\label{e5}	
\end{eqnarray}
and
\begin{eqnarray}
\! \! \! y_n \! \! &=& \! \! \pm 2 \sqrt{P_2} \cos[\Phi_2 + (n-1)2 \pi/3], 
                             \> (n=1,2,3), \\
\! \! \! P_2 \! \! &=& \! \! [(J_{AF} - J_2^{+})/6]^2 + 
                             [(J_1^{-})^2 + 2 (J_{AF} \Delta)^2]/12,  \\
\! \! \! Q_2 \! \! &=& \! \! [(J_{AF} - J_2^{+})/6]^3 + \nonumber \\ 
&& \quad + [(J_{AF} - J_2^{+})/6]  [(J_{AF} \Delta)^2 - (J_1^{-})^2]/4,  \\
\! \! \! \Phi_2 \! \! &=& \! \! 
         \frac13 \arctan\Bigl( \sqrt{P_2^3 - Q_2^2}/Q_2 \Bigr).
\label{e6}	
\end{eqnarray}
Note that the signs of $x_n$ and $y_n$ are unambiguously determined by 
the signs of the expressions $Q_1$ and $Q_2$, respectively.

It should be emphasized that the mapping parameters $A$, $R$ and $H_0$ 
are "self-consistently" given by the transformation Eq. (\ref{r4}), 
which must be valid for any combination of spin states of the $S_{4k-4}$ and $S_{4k}$ Ising spins. In consequence of that, the mapping parameters can be 
obtained following the standard procedure \cite{DIT, LC, DC, 2D} from the expressions:
\begin{eqnarray}
A^4 = V_1 V_2 V_3^2, \hspace*{2mm} \beta R = \ln(V_1 V_2/ V_3^2), \hspace*{2mm}
\beta H_0 = \ln(V_1/V_2), \nonumber \\
\label{e7}	
\end{eqnarray}
where the functions $V_1$, $V_2$ and $V_3$ have a physical meaning of 
the bond partition function (\ref{r4}), when taking into account all the particular 
spin combinations of the $S_{4k-4}$ and $S_{4k}$ Ising spins. These, however,  modify merely the effective coupling constants (\ref{e4})-(\ref{eee4}) implicitly contained in Eq. (\ref{r4}). Thus, the functions $V_1$, $V_2$, and $V_3$ are definitely determined by this set of expressions:
\begin{eqnarray}
V_1 = Z_k, \hspace*{0.75mm} \mbox{if:} \hspace*{0.5mm}
J_1^{+} \! \! &=& \! \! J_F + 2B_2 + B_3, \hspace*{6.5mm} J_1^{-} = 0,  \nonumber \\
J_2^{+} \! \! &=& \! \!  J_F + 2(B_2 - B_3);   \\
V_2 = Z_k, \hspace*{0.75mm} \mbox{if:} \hspace*{0.5mm}
J_1^{+} \! \! &=& \! \! -J_F + 2B_2 + B_3, \hspace*{3.5mm} J_1^{-} = 0, \nonumber \\
J_2^{+} \! \! &=& \! \! -J_F + 2(B_2 - B_3);   \\
V_3 = Z_k, \hspace*{0.75mm} \mbox{if:} \hspace*{0.5mm}
J_1^{+} \! \! &=& \! \! 2B_2 + B_3, \hspace*{15mm} J_1^{-} = J_F, \nonumber \\
J_2^{+} \! \! &=& \! \! 2(B_2 - B_3).       
\label{e8}	
\end{eqnarray}

When substituting Eq. (\ref{r4}) into Eq. (\ref{r3}), the transformation Eq. (\ref{r4}) maps the original F-F-AF-AF Ising-Heisenberg BAC on the uniform spin-1/2 Ising chain with the effective nearest-neighbor exchange coupling $R$ and the magnetic field $H_0$. As a result, the partition function $Z$ of the Ising-Heisenberg BAC can be directly related to the partition function $Z_{0}$ of the corresponding Ising chain:
\begin{eqnarray}
Z = A^{N} Z_{0} (\beta, R, H_0).
\label{r5}	
\end{eqnarray}
Certainly, similar mapping relations can be established also for other thermodynamic quantities. For instance, the Gibbs free energy $G$ of the Ising-Heisenberg BAC can be calculated from the relevant expression of the Gibbs free energy $G_0$ of the corresponding spin-1/2 Ising chain: 
\begin{eqnarray}
G = G_0 (\beta, R, H_0) - N k_B T \ln(A).
\label{r6}	
\end{eqnarray}
Since an exact solution for the spin-1/2 Ising chain was known a long time ago, 
\cite{IC} the above equation can serve as the basic generating equation 
from which all thermodynamic quantities can be extracted. 
Here, we shall restrict ourselves to the analysis of the magnetization, susceptibility and specific heat. For the spin only value of the on-site magnetization 
at the magnetically nonequivalent Cu1, Cu2 and Cu3 positions, we shall introduce 
this simple notation: 
\begin{eqnarray}
m_1 = \langle \hat{S}_{4k}^{z} \rangle,  \quad
m_2 = \langle \hat{S}_{4k-1}^{z} \rangle, \quad
m_3 = \langle \hat{S}_{4k-2}^{z} \rangle, 
\label{r7}	
\end{eqnarray}
where the symbol $\langle ... \rangle$ stands for a standard canonical average over the ensemble defined by the Hamiltonian (\ref{r1}). In view of this notation, the total magnetization normalized per one Cu$^{2+}$ ion can be expressed as follows: 
$m = \mu_B (g_1 m_1 + 2 g_2 m_2 + g_3 m_3)/4$.

Finally, we briefly mention the basic thermodynamic relations from which all the analyzed quantities have been calculated after straightforward but a little bit lengthly calculations. The on-site magnetization can be obtained by differentiating the Gibbs free energy with respect to the particular magnetic fields:
\begin{eqnarray}
m_1 = - \frac{1}{N} \frac{\partial G}{\partial B_1},
m_2 = - \frac{1}{2N} \frac{\partial G}{\partial B_2}, 
m_3 = - \frac{1}{N} \frac{\partial G}{\partial B_3}, 
\label{r9}	
\end{eqnarray}
while the susceptibility and specific heat have been obtained
as the second derivatives of the Gibbs potential using the standard 
thermodynamic relations:
\begin{eqnarray}
\chi =  - \frac{\partial^2 G}{\partial B^2}, \qquad \mbox{and} \qquad
C = - T \frac{\partial^2 G}{\partial T^2}. 
\label{r10}	
\end{eqnarray}
It should be stressed that the explicit form for these quantities is too 
cumbersome to write it down here; however, it can be obtained from the 
authors on request.

\section{Results and discussion}

Before proceeding to a discussion of the most interesting results,
the model reliabi\-lity should be checked in connection with its 
possible application to interpret the experimental data on CCPA,
because a danger of overinterpretation is inherent in any approximation.
At first glance, we have made in our model a very crude conjecture in that 
the F interaction $J_F$ was approximated by an Ising-type coupling although 
all Cu$^{2+}$ ions are nearly isotropic, and whence, the Heisenberg interaction 
would be more appropriate. With regard to this, Fig. 2 illustrates a comparison between the on-site magnetization of the Ising-Heisenberg BAC and that 
of the corresponding pure Heisenberg BAC obtained from the exact diagonalization method for a finite-size cluster of 12 sites. \cite{Hagiwara2} 
It should be pointed out that the low-temperature ($k_B T / J_{AF} = 0.05$) magnetization curve provides the best independent test of the model reliability, 
since it reflects magnetization near the ground state, where the clearest mani\-festation of the quantum fluctuations should be expected to occur. 

It is quite surprising that there is such an excellent agreement between the two theories: the total magnetization (for clarity not shown here, but see for instance Fig. 4) exhibits a steep increase from  zero field followed by a magnetization plateau and finally, there appears a second steep increase near the transition field toward a fully polarized state. Strictly speaking, there is no real phase transition at any finite temperature; however, the low-temperature magnetization curve shed light on what happens in the ground state: the plateau state should reflect the ground-state phase and the field-induced transition to the fully saturated phase takes indeed place in the zero-temperature limit. Moreover, the magnetic order at the plateau state has a typical feature of quantum ferrimagnet, in fact, one finds here a substantial quantum reduction of the magnetization $m_2$ and $m_3$ at the Cu2 and Cu3 sites even though the magnetization $m_1$ (Cu1 site) retains its saturation value. It means, among other matters, that the spin deviations cannot propa\-gate through these sites and hence, they are strictly localized within the AF trimers. This explains also the remarkable agreement between both the theories, since the quantum fluctuations are in our model artificially restricted to the AF trimers because of the presence of the Ising spins. 

The most obvious difference between the magnetization curve of the Ising-Heisenberg BAC and the pure Heisenberg BAC can be thus found in the vicinity of the zero field, where the on-site magnetization of the former reaches their plateau values more rapidly. Apparently, this distinction can be attributed to the more "susceptible" character of the Ising spins, whereby their Heisenberg counterparts achieve the plateau values more rapidly on behalf of the local field produced just by the Ising spins. With this in mind, small systematic discrepancies should be expected to occur especially in the region where both $B \to 0$ and $T \to 0$, but, without loss of the qualitative agreement. 

Why the Cu1 sites represent a "barrier" for the spin deviations also in the spin-1/2 Heisenberg F-F-AF-AF BAC still remains an open question. Although the origin of this outstanding feature could be naively understood as a general feature of the spins coupled by the F interactions only, this suggestion is 
in an obvious contradiction with Hida's results for the spin-1/2 F-F-AF BAC. \cite{FFAFtheor} As a matter of fact, the total magnetization of this system varies smoothly with increasing  magnetic field even in the ground state, which implies that the spin deviations are delocalized over the whole chain and thus, there is a certain quantum reduction also at purely ferromagnetically coupled sites. 

\subsection{Survey of theoretical results}

In this part, an extensive survey of theoretical results will be presented
in order to enable deep insight into the magnetic behavior of the system 
under investigation. For the sake of simplicity, we shall firstly suppose 
equal $g$-factors at the Cu1, Cu2, and Cu3 positions, i. e. 
$g_1 = g_2 = g_3 = g$. 

We start our discussion with the analysis of the ground state. Below some critical external magnetic field, the ground state exhibits an interesting ferri\-magnetic order, which can be characterized by the following values of the ground-state magnetization: 
\begin{eqnarray}
m_1 \! \! &=& \! \! \frac12, \quad 
m_2 = \frac14 \Bigl[ 1 + \frac{1 + J_F/J_{AF}}
                           {\sqrt{(1 + J_F/J_{AF})^2 + 8 \Delta^2}} \Bigr], 
\nonumber \\
m_3 \! \! &=& \! \! - \frac12 \frac{1 + J_F/J_{AF}}
                     {\sqrt{(1 + J_F/J_{AF})^2 + 8 \Delta^2}}, \quad
\frac{m}{m_s} = \frac{1}{2},
\label{r11}	
\end{eqnarray}
where $m_s$ labels the saturation magnetization normalized per one Cu$^{2+}$ ion.
As already mentioned in the preceding part, the magnetization $m_1$ retains 
in the ground state its saturation value in contrast to the magnetization $m_2$ 
and $m_3$, which show obvious quantum reduction. It is quite evident that the stronger the exchange anisotropy $\Delta$, the greater the reduction of the magnetization $m_2$ and $m_3$. On the other hand, the quantum reduction of the magnetization completely vanishes in the Ising limit $\Delta \to 0$, when the ground state exhibits a "classical" ferrimagnetic order $\uparrow \uparrow \downarrow \uparrow \dots$ (Cu1Cu2Cu3Cu2...) of a rather trivial nature. 

As expected, the ferrimagnetic system undergoes a field-induced metamagnetic  transition toward the fully saturated phase under the certain magnetic field. 
It can be readily proved that the transition field $B_t$ is given under the condition:
\begin{eqnarray}
g \mu_B B_t / J_{AF} = \frac14 \Bigl(3 \! \! \! &-& \! \! \!  J_F/J_{AF} + 
\nonumber \\ \! \! \! &+& \! \! \!  \sqrt{(1 + J_F/J_{AF})^2 + 8 \Delta^2} \Bigr).
\label{r12}	
\end{eqnarray}
For illustration, Fig. 3 displays the ground-state phase diagram in the 
$J_F - B$ plane for several values of the exchange anisotropy $\Delta$. 
As one can see, increasing strength of the F coupling $J_F$ generally reinforces the gradual decline of the transition field. Thereby, the pure Ising limit 
($\Delta = 0$) represents the only exceptional case when the magnitude of the transition field does not change as the ratio $J_F/J_{AF}$ varies. It can be easily understood that the metamagnetic transition arises in this particular case from a single flip of the central (Cu3) spin of each AF trimer, which occurs when the exchange energy of $J_{AF}$ is thoroughly balanced by the external magnetic field ($g \mu_B B_t = J_{AF}$). 

The situation becomes much more complex on considering the nonzero exchange anisotropy $\Delta$. Due to the quantum fluctuations, the spin oriented in an opposite direction with respect to the external field is no longer rigidly connected with the central spin of the AF trimer, but is collectively held by the entire trimer. In other words, the reversed spin is delocalized over the whole AF trimer and consequently there is a non-zero probability that the side (Cu2) spins of the  trimer are aligned opposite to the field direction. Natu\-rally, this must lead to the enhancement of the exchange energy between the ferromagnetically coupled Cu2 side spins and the fully polarized Cu1 spins. In consequence of that, it is quite conspicuous that the suppression of the transition field can be explained in terms of the energetic destabilization of the quantum ferrimagnetic order, which occurs when $J_F/J_{AF}$ increases.

Now, let us turn our attention to the magnetic behavior at finite tempera\-tures. 
In Fig. 4 we plot the total magnetization scaled in  $g \mu_B$ units against 
the dimensionless magnetic field ($g \mu_B B / J_{AF}$) for a few temperatures. 
It should be pointed out that the magnetization curve starts at any finite temperature from zero according to the one-dimensional character of the spin 
system. The sharp stepwise magnetization curve observable in the zero-temperature limit is, however, gradually smeared out as the temperature is raised from zero. Actually, the conversion toward the fully saturated phase does not occur merely at one precise value of the transition field, but is smudged over a finite range of the fields. In addition, the temperature-induced fluctuations greatly shrink also 
the width of the magnetization plateau and above a certain temperature, the plateau state completely disappears from the magnetization curve. By any means, Fig. 4 provides further convincing evidence that the observed magnetization plateau emerges, under the assumption of the uniform $g$-factors, exactly at one-half of the saturation magnetization [see also Eq. (\ref{r11})].

The variation of the field-induced magnetization with the temperature (Fig.~5) 
is also interesting on account of the magnetically ordered ground state. 
At low magnetic fields, the total magnetization tends abruptly to zero as the temperature increases. Nevertheless, one finds here also a noticeable temperature-induced increase of the magnetization within a range of moderate fields,
when the total magnetization shows a broad maximum resulting from a vigorous thermal excitations of the Cu3 spins and smaller thermal excitations of the Cu2 spins. However, when the magnetic field exceeds the transition field given by the condition (\ref{r12}), the total magnetization monotonically decreases with increasing temperature as it already starts from its maximum value.

The thermal dependence of the zero-field susceptibility times temperature 
($\chi T$) data is displayed in Fig. 6. Interestingly, the $\chi T$ product 
exhibits a round minimum upon cooling and then diverges under further 
tempe\-rature suppression. As already discussed in several papers concerned 
with the QFC, \cite{Yama} a thermal dependence of this type reveals a ferromagnetic-to-antiferromagnetic crossover. Really, the temperature variation 
of the $\chi T$ data is in general a monotonically decreasing function for ferromagnets, but monotonically increasing for antiferromagnets. 
In this respect, the marked low-temperature divergence of the susceptibility emerges markedly because of the gapless F excitations from the magnetically ordered ground state. The position of the round minimum, on the other hand, designates the temperature above which the gapped excitations of the AF nature overhelm 
the gapless excitations originating from $J_F$. In accordance with the above statement, the round minimum flattens as the ratio $J_F/J_{AF}$ strengthens and simultaneously its position is shifted toward higher temperatures.

As far as the magnetic susceptibility at nonzero fields is concerned (Fig.~7),
the $\chi T$ always initiates from zero because of the energy gap opened by the magnetic field. However, the thermal dependence of the $\chi T$ data shows at weak fields a dramatic increase until it reaches a sharp maximum, which is followed by the familiar round minimum of the same origin, as already discussed by the zero-field susceptibility. The appearance of the additional sharp peak can obviously be related to a thermal instabi\-lity of the magnetically ordered system, because already a small temperature change necessitates a huge variation of the magnetization at low fields (see Fig. 5). As a matter of fact, 
the $\chi T$ product rises steadily with the temperature at sufficiently strong fields, since the thermal fluctuations are not strong enough to induce a simultaneous excitation of a large number of spins. Thus, an interesting thermal dependence of the susceptibility appears for strong magnetic fields just around 
the transition to the fully polarized state. Under these circumstances, the $T$-$\chi T$ plot exhi\-bits a rapid increase over a small temperature range, which is subsequently followed by a narrow plateau that continuously passes into a slowly repeating increase of the susceptibility (see the curve $g \mu_B B/J_{AF} = 1.4$). 

To receive a complete picture of the thermodynamics, the thermal variations of the specific heat are plotted in Figs. 8 and 9. When considering the zero-field case (Fig. 8), the specific heat displays as a function of the temperature a remarkable double-peak structure. There are strong indications that the first sharper peak observed at lower temperature originates from the F excitations: the peak becomes wider as the ratio $J_F/J_{AF}$ increases and is shifted to higher temperatures notwithstanding its almost constant height. 
Under the assumption of dominant F coupling constant ($J_F/J_{AF} > 1.0$), the low-temperature peak 
is to a large extent superimposed on the second broader maximum that occurs at a little bit higher temperature. Obviously, the round high-temperature maximum can be thought of as the usual Schottky-type maximum, which has a tendency to be enhanced in magnitude with increasing $J_F/J_{AF}$. When comparing the results displayed in Fig. 8 with those of the pure Heisenberg F-F-AF-AF BAC obtained using the exact diagonalization method \cite{Hagiwara3} or the transfer-matrix renormalization group \cite{Lu}, an excellent agreement is found as far as the high-temperature maxima of both these models are concerned. The only difference 
thus rests in the height of the low-temperature peak which is, due to the Ising approximation of $J_F$, 
roughly three-times higher for the Ising-Heisenberg BAC compared to that of the pure Heisenberg 
BAC (see for instance the insert of Fig. 4 depicted in Ref. \onlinecite{Lu}). 

The situation becomes even much more interesting on applying a magnetic field (Fig. 9). The effect of a small magnetic field is to increase the height of the low-temperature peak and to move it toward higher temperatures (see the case $g \mu_B B / J_{AF} = 0.1$). This result is taken to mean that under a certain 
field both maxima coalesce and consequently, the specific heat exhibits a single nonrounded maximum as shown for $g \mu_B B / J_{AF} = 0.2$. By an additional increase of the field strength, the specific heat curve gradually loses its irregular profile and the overall trend is that the Schottky-type maximum drops in magnitude and moves to higher temperatures. Apart from this rather trivial finding, the double-peak specific heat curve can be recovered for the fields close to the transition to the fully saturated phase (e.g. $g \mu_B B / J_{AF} = 1.4$). It should be mentioned, however, that the height of the low-temperature peak is in this case considerably smaller than that of the zero-field specific heat. Although the external field spreads also this low-temperature peak until it completely merges with the Schottky-type maximum, the change of the peak position occurs strikingly without any significant change of the peak height (for clarity, this effect is not shown here). Finally, it should be stressed that the observed behavior of the specific heat is not a generic feature of the special class of the Ising-Heisenberg BAC, nevertheless, it has already been recognized also in the pure Ising BAC (compare with Figs. 4 and 8 from Ref. \onlinecite{Minami}).

We conclude our survey of the thermodynamic pro\-perties 
by considering the nonuniformity of the $g$-factors at the structurally and therefore also magneti\-cally nonequivalent Cu1, Cu2 and Cu3 positions (i.e. $g_1 \neq g_2 \neq g_3$). Let us firstly focus on the ground-state behavior. The most interesting finding to emerge here is that the ground-state magnetization depend on the field strength, whenever $g_2 \neq g_3$:
\begin{widetext}
\begin{eqnarray}
m_1 \! \! &=& \! \! \frac12, \quad 
m_2 = \frac14 \biggl \{ 1 + \frac{1 + J_F/J_{AF} + 2 \mu_B B (g_2 - g_3)/J_{AF}}
 {\sqrt{[1 + J_F/J_{AF} + 2 \mu_B B (g_2 - g_3)/J_{AF}]^2 + 8 \Delta^2}} \biggr \}, 
\nonumber \\
m_3 \! \! &=& \! \! - \frac12 \frac{1 + J_F/J_{AF} + 2 \mu_B B (g_2 - g_3)/J_{AF}}
   {\sqrt{[1 + J_F/J_{AF} + 2 \mu_B B (g_2 - g_3)/J_{AF}]^2 + 8 \Delta^2}}, \nonumber \\
\frac{m}{m_s} &=& \frac{g_1 + g_2}{g_1 + 2 g_2 + g_3} 
               + \frac{g_2 - g_3}{g_1 + 2 g_2 + g_3} \frac{1 + J_F/J_{AF} + 2 \mu_B B (g_2 - g_3)/J_{AF}}{\sqrt{[1 + J_F/J_{AF} + 2 \mu_B B (g_2 - g_3)/J_{AF}]^2 + 8 \Delta^2}}.
\label{r13}	
\end{eqnarray}
\end{widetext}
It is quite apparent from Eqs. (\ref{r13}) that the field dependence of the total magnetization comes from the corresponding field variations of the on-site magnetization $m_2$ and $m_3$. Another noticeable feature to observe here is that the magnetic field suppresses (raises) the quantum reduction of the magnetization $m_2$ and $m_3$ as far as $g_2 > g_3$ ($g_2 < g_3$). Since the condition $2 \mu_B B (g_2 - g_3) \ll J_F + J_{AF}$ holds for most of the experimentally accessible fields and $g$-factors,  the total magnetization should be nearly linearly dependent on the magnetic field with a linear term proportional to the diffe\-rence $\delta g = g_2 - g_3$. Naturally, the greater the magnitude of the external field and $\delta g$, the stronger deviations from the linearity should emerge. 

Some typical examples of the ground-state magnetization curves are depicted in 
Fig. 10 under the assumption $g_2>g_1=g_3$ (Fig. 10a) and $g_3>g_1=g_2$ 
(Fig.10b). These results can serve as evidence that the magnetization curve does not show an exact plateau, but rises steadily with the external field. It is also worthy to note that these "magnetization plateaus with a finite slope" (when plotted in the full range with respect to the satu\-ration magnetization, they are  hardly discernible from the exact plateaus within the reasonable values of the $g$-factors) do not occur precisely at half of the saturation magnetization ($m/m_s=1/2$), but they are shifted to higher values if $g_2>g_1=g_3$ (Fig. 10a) and respectively, to the smaller values if $g_3>g_1=g_2$ (Fig. 10b). 

To enable an independent check of the field dependence of the total magnetization, the temperature variation of the susceptibility has been examined in detail. 
The typical thermal dependences of the susceptibility are plotted in Fig. 11 
for $g_3 > g_1 = g_2$ and some selected external fields. Generally, the  susceptibi\-lity versus temperature plot can be characterized by a round maxi\-mum, which flattens and shifts to higher temperatures when increasing 
the external field. The low-temperature part of the susceptibility is displayed 
on an enlarged scale in the inset. It turns out that the susceptibility 
does not completely vanish as the temperature approaches zero; 
nevertheless, it tends toward a small but finite value. This result provides 
a confirmation of the striking field-dependent magnetization and moreover, 
it also clearly demonstrates a gapless excitation spectrum above the ground state. 

\subsection{Comparison with the experimental data}

At this stage, the results obtained from the exact mapping on the Ising-Heisenberg BAC will be compared with the available experimental data for a single-crystal sample of the CCPA. It should be mentioned here that all the experimental data used in our further analysis have already been reported in earlier publications \cite{Hagiwara2,Hagiwara3} to which the interested reader is referred for more expe\-rimental details. 

In order to fit the experimental data of the CCPA, we have to consider the six fitting parameters: the exchange constant $J_{AF}/k_B$, the exchange anisotropy $\Delta$, the alternation ratio $J_F/J_{AF}$ and the three $g$-factors $g_1$, $g_2$ 
and $g_3$. Note that the best fit of the magnetization and the susceptibility data obtained under the restriction of equal $g$-factors ($g_1=g_2=g_3$) as well as isotropic exchange $J_{AF}$ ($\Delta = 1.0$), have already been published by the present authors in our preliminary report (see Fig. 3 in Ref. \onlinecite{ccpa}). 
Unless specifically mentioned, the exchange anisotropy will again be fixed 
to the purely isotropic case $\Delta = 1.0$, because the experimental measurements on the single-crystal sample do not reveal any significant spatial dependence: 
the spatial directions parallel and perpendicular to the chain axis seem to be almost equivalent. \cite{Hagiwara2} 

In Fig. 12, the low-temperature magnetization curve of CCPA measured 
in the pulsed (Fig. 12a) and static (Fig. 12b) magnetic fields is compared with the relevant theoretical prediction (the fitting parameters are indicated in the figures). It turns out, however, that the average $g$-factor must be approximately 
$g_{av} \equiv (g_1 + 2 g_2 + g_3)/4 \sim 2.15$ in order to get the correct satu\-ration magnetization (Fig. 12a). Even under this confinement, several combinations of the $g$-factors provide almost the same fit to the experimental magnetization curve. For illustrative purposes, we choose the one with the smallest value of the $g_1$-factor to lower the susceptibility of the Cu1 spins, because the most obvious discrepancies between the theory and the experiment emerge near  zero field due to the more "susceptible" character of the Ising spins, as already discussed earlier.

Unfortunately, it is beyond experimental verification to find out from the existing data set, whether the magnetization increase observed within the plateau region appears exclusively on account of the finite-temperature effect, or is partially caused by the inequa\-lity $g_2 \neq g_3$. Although the former situation cannot be definitely ruled out, it is rather more acceptable that there is at least a small contribution to the magnetization increase also from the difference between the $g$-factors. Nonetheless, the final conclusion is complicated by different factors: the magnetization curves in the static fields (Fig. 12b) are measured at higher temperatures and become quite noisy above 15 T, while the available magnetization curve in the pulsed fields (Fig. 12a) show a relatively large hysteresis probably caused by the magnetocaloric effect. \cite{Hagiwara2} 

The thermal dependence of the zero-field susceptibility times temperature 
data is illustrated in Fig.~13. To obtain a reasonable accordance with the experimental measurement, slightly higher $g$-factors must be taken into consideration in comparison with the ones acquired by the fitting of the magnetization. Despite this fact, a quite impressive accordance can be achieved by rescaling the average $g$-factor at about 4-5\% of its magnetization value $g_{av} \sim 2.15$. As expected, the most appa\-rent disagreement appears in the descending tail of the $\chi T$ curve nearby the zero temperature, where the susceptibility divergence is reminiscent of the ferromagnetic Ising-type divergence rather than the Heisenberg-type divergence.

Finally, Fig. 14 shows the magnetic part of the specific heat as a function of 
the temperature for several magnetic fields. The corresponding theoretical predictions are also included; nevertheless, the specific heat can merely be fitted with a rather unacceptable precision even though the results agree at least qualitatively. In order to situate the striking low-temperature peak around $0.5$ K, however, a drastic reduction of the $J_F$ and $J_{AF}$ exchange parameters was carried out. Even under this condition, the predicted peak is much more robust than observed experimentally and this feature remains valid also for any other combination of the $J_F$ and $J_{AF}$ coupling constants. Actually, the height of 
the low-temperature peak does not significantly vary with the ratio $J_{F}/J_{AF}$ as already shown in Fig. 8. 

It is quite obvious that the inconsistency in the fitting set for the specific heat and, respectively, the magnetization and susceptibility, indicates some insufficiency in the considered model based on the simple concept of the F-F-AF-AF BAC. Notice that the basic problem that prevents obtaining an unique fitting set 
for all these quantities consists in reproducing the striking low-temperature peak of the specific heat, 
which cannot be fitted within a reasonable accord neither under the assumption of the pure Heisenberg 
F-F-AF-AF BAC.\cite{Hagiwara3, Lu} A possible explanation for this peculiar inconsistency has recently 
been suggested by Lu {\it et al.} \cite{Lu}, who found a strong indication that the striking low-temperature peak comes from the contribution of magnetic impurities. Until a higher-quality sample 
of CCPA is prepared and remeasured, however, one cannot exclude the other possibility that this anomalous peak originates from the neglected higher-order interaction terms in the proposed Hamiltonian. Our detailed analysis shows that neither the nonuniformity of $g$-factors nor the exchange anisotropy of the 
$J_{AF} (\Delta)$ coupling constant, can resolve this inconsistency.

\section{Concluding remarks}

In the present article, the magnetic properties of the spin-1/2 Ising-Heisenberg chain with  regular F-F-AF-AF bond alternation have been investigated within the exact mapping transformation method. The results 
obtained from the mapping procedure are compared with the results of the corresponding Heisenberg BAC \cite{Hagiwara2} and with the experimental data of the CCPA compound, \cite{Hagiwara2, Hagiwara3} which is regarded to be an experimental realization of the spin-1/2 F-F-AF-AF BAC.

It should be emphasized that all characteristic quantum features observed in the spin-1/2 Heisenberg F-F-AF-AF BAC \cite{Hagiwara1, Hagiwara2, Hagiwara3} are still present also in our simplified Ising-Heisenberg version: one finds here a substantial quantum reduction of the ground-state magnetization $m_2$ and $m_3$ at the Cu2 and Cu3 sites in addition to the fully saturated ground-state magnetization $m_1$ at the Cu1 site. Perhaps, the most striking finding stemming from our study is that there appears a rather unusual change from the gapful to gapless excitation spectrum when converting the uniform $g$-factors at  structurally non-equivalent positions to  nonuniform ones. With regard to this, the magnetization curve exhibits a real plateau under the condition $g_2 = g_3$ only; for any other case the magnetization within the quantum ferrimagnetic phase (plateau state) shows a weak field dependence. 

The success of the simplified Ising-Heisenberg model in reproducing the measured data for CCPA is also quite remarkable. With the exception of the specific heat, where the results agree at the qualitative level only, other thermodynamic quantities are relatively well reproduced. In fact, small systematic errors occur merely in the low-temperature and weak-field region, where the approximation of the ferromagnetic coupling $J_F$ by the Ising-type interaction plays the most essential role. It is worthwhile to mention, moreover, that the discrepancy found in the specific heat fit cannot be resolved even if a pure Heisenberg F-F-AF-AF BAC 
is considered. Actually, very recent work of Lu {\it et al.}\cite{Lu} provided a strong indication 
that the striking low-temperature peak of specific heat is being of extrinsic origin probably caused by 
the presence of magnetic impurities. However, further experimental measurements with high-quality samples 
of CCPA are required to clarify this issue. Finally, one should notice that the simple concept of the Ising-Heisenberg F-F-AF-AF BAC presented here can be rather straightforwardly extended to account for second-neighbor interactions, the antisymetric Dzyaloshinsky-Moriya interaction, and/or multispin interactions, which cannot be ruled out as a potential cause of the observed discrepancies until 
the influence of magnetic impurities will experimentally be confirmed.  

\begin{acknowledgements}
One of the authors (J. S.) would like to thank Dr. M.~Jur\v{c}i\v{s}in and 
Prof. N.~M.~Plakida for a kind hospitality during his stay in the Joint Institute 
for Nuclear Research (JINR) in Dubna, where the part of this work was comple\-ted.
The author is also indebted to Prof. J.~Richter and Dr. V. N. Plechko for 
fruitful discussions and valuable suggestions. 

One of the authors (M.H.) express his sincere thanks to Dr. H. A. Katori, 
Dr. H. Kitazawa, Dr. H. Suzuki, Dr. N. Tsujii and Dr. H. Abe for their help 
in doing the experiments. Some of the experimental works were done under 
the Visiting Researcher Program of KYOKUGEN at Osaka University.

This work was partially (J. S. and M. J.) supported under the grants VEGA 1/2009/05 and the APVT-20-005204.

\end{acknowledgements}

\vspace*{0.75cm}

\begin{large}
\textbf{Figure captions}
\end{large}

\begin{itemize}

\item [Fig. 1]
The fragment of the CCPA structure. Fig. 1a shows a full details with the azido (N$_3$) bridges as well as the bulky 3-Clpy ligands, Fig. 1b schematically reproduces the considered magnetic structure only. There are three non-equivalent positions of the Cu$^{2+}$ ions: the Cu1 sites coupled purely by the F interaction (black circles), the Cu2 sites coupled via both F and AF interactions (grey circles) and lastly, the Cu3 sites coupled purely by the AF interaction (light-grey circles). The rectangles designate the AF trimers.

\item [Fig. 2]
The comparison between the on-site magnetization of the Ising-Heisenberg BAC and 
the corresponding pure Heisenberg BAC obtained by the exact diagonalization method as described in Ref. \onlinecite{Hagiwara2}. For details see the text.

\item [Fig. 3]
The ground-state phase diagram in the $J_F$-$B$ plane obtained for several exchange anisotropies $\Delta$ under the assumption of the uniform $g$-factors, 
i.~e. $g_1 = g_2 = g_3 = g$.

\item [Fig. 4]
The total magnetization scaled in the $g \mu_B$ units as a function of the dimensionless magnetic field $g \mu_B B / J_{AF}$ for $J_F / J_{AF} = 0.5$,
$\Delta = 1.0$ and several dimensionless temperatures $k_B T / J_{AF}$.

\item [Fig. 5]
Some typical temperature variation of the total magnetization obtained for 
$J_F / J_{AF} = 0.5$, $\Delta = 1.0$ and several magnetic fields.

\item [Fig. 6]
The thermal dependence of the zero-field susceptibility times temperature data scaled in the $\mbox{N}_{{\rm A}} (g \mu_B)^2 / k_B$ units ($\mbox{N}_{{\rm A}}$ - Avogadro's number) for $\Delta = 1.0$ and some typical values of the $J_F/J_{AF}$.

\item [Fig. 7]
The temperature dependence of the $\chi T$ product as a function of the field strength when $J_F/J_{AF} = 0.5$ and $\Delta = 1.0$.

\item [Fig. 8]
The zero-field specific heat as a function of the temperature for some typical values of the $J_F/J_{AF}$ and $\Delta = 1.0$. The specific heat is scaled 
in multipliers of the universal gas constant R = 8.314 J.K$^{-1}$.mol$^{-1}$.

\item [Fig. 9]
The thermal dependence of the specific heat by various magnitudes of the 
external magnetic field when $J_F/J_{AF} = 0.5$ and $\Delta = 1.0$.

\item [Fig. 10]
The field variation of the ground-state magnetization normalized with respect 
to its saturation value by assuming the non-uniformity of the $g$-factors. 
Fig. 10a (Fig. 10b) shows the situation when $g_2 > g_1 = g_3$ ($g_3 > g_1 = g_2$).

\item [Fig. 11]
The temperature dependence of the susceptibility when varying the external field strength, and $J_F/J_{AF} = 0.5$, $\Delta = 1.0$, $g_1 = g_2 = 2.0$ and $g_3 = 2.25$. The insert displays the low-temperature susceptibility in enlargened scala. 

\item [Fig. 12]
The low-temperature magnetization curves measured in the pulsed (Fig.~12a) 
and static (Fig.~12b) magnetic fields together with the corresponding theo\-retical prediction obtained for this fitting set of parameters: $J_{AF}/k_B=28.2$ K, $J_F/J_{AF} = 0.4$, $\Delta = 1.0$, $g_1 = 2.10$, $g_2 = 2.14$ and $g_3 = 2.19$. The hysteresis observed in the experimental curves near the zero and saturation 
fields is probably caused by the magnetocaloric effect.

\item [Fig. 13]
The zero-field susceptibility times temperature data and the corresponding theoretical prediction (the fitting parameters are indicated in the figure).

\item [Fig. 14]
The thermal dependence of the specific heat measured at $B = 0.0$, $0.5$ and $1.0$ T.
The corresponding theoretical predictions are also included. 

\end{itemize}

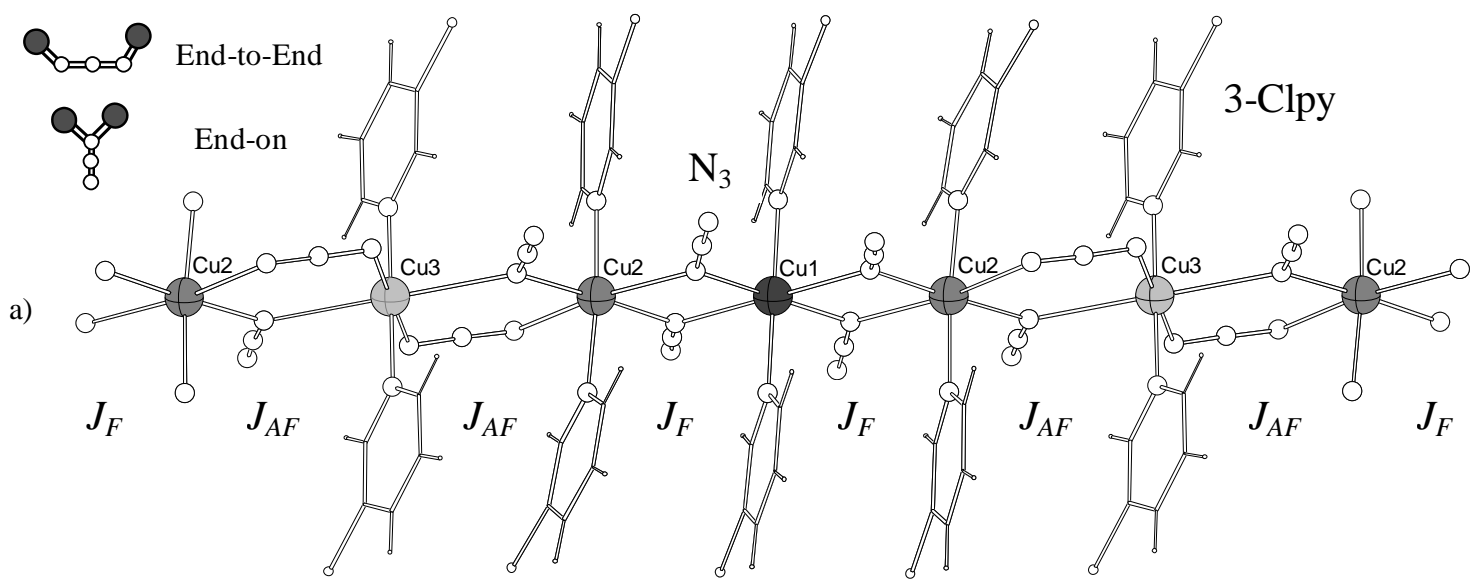
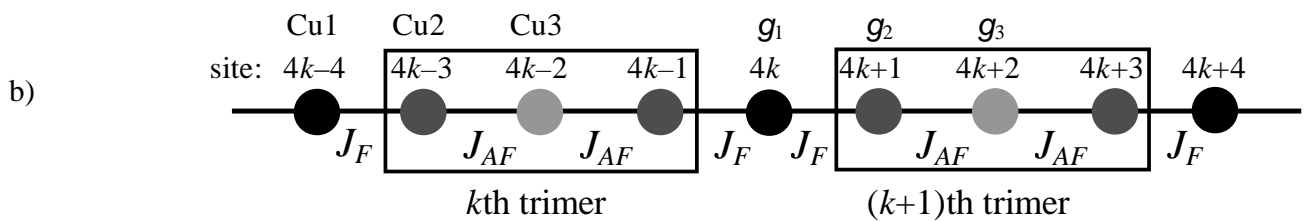
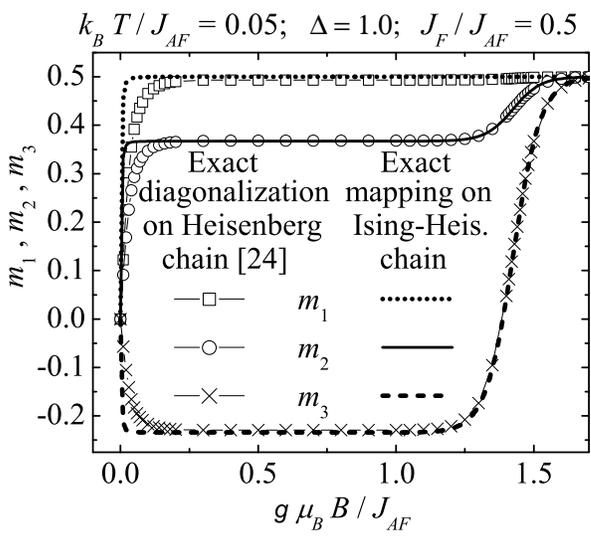
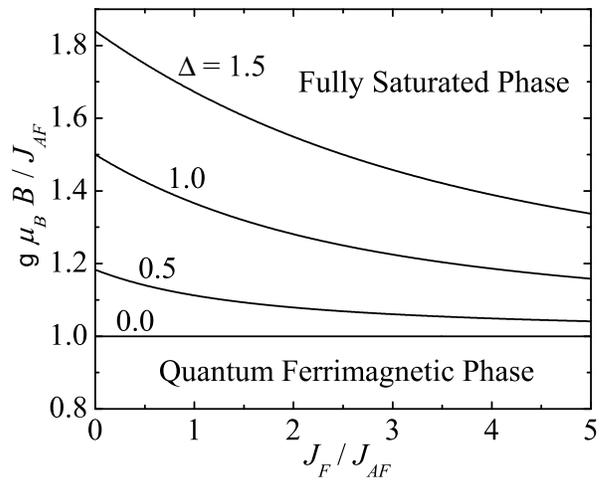
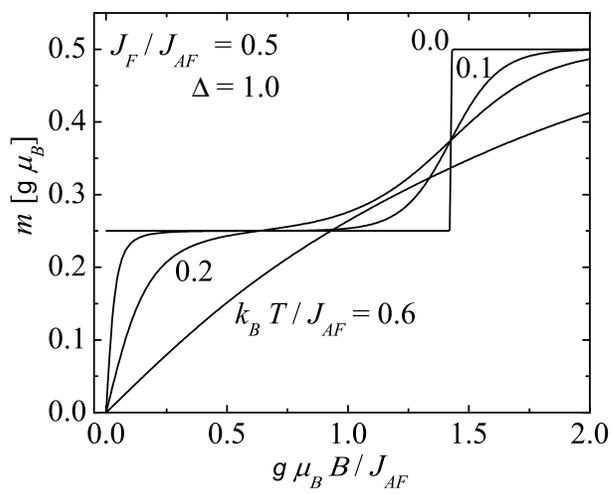
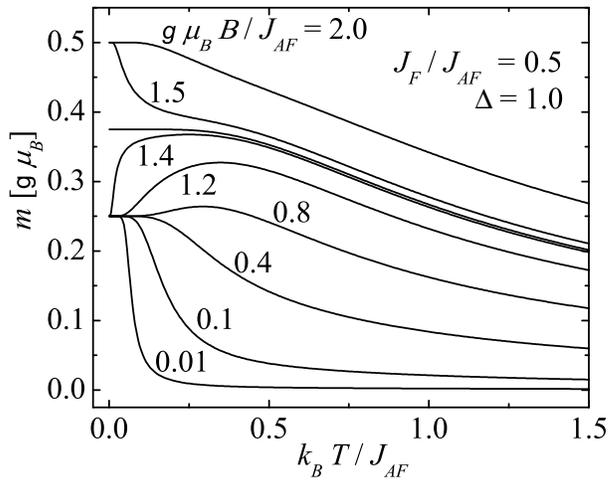
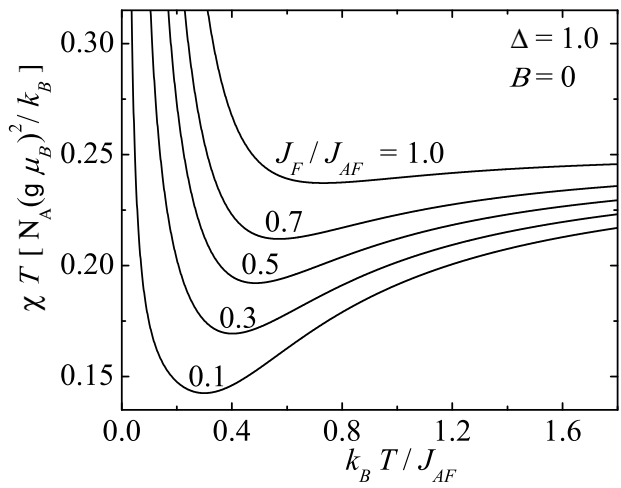
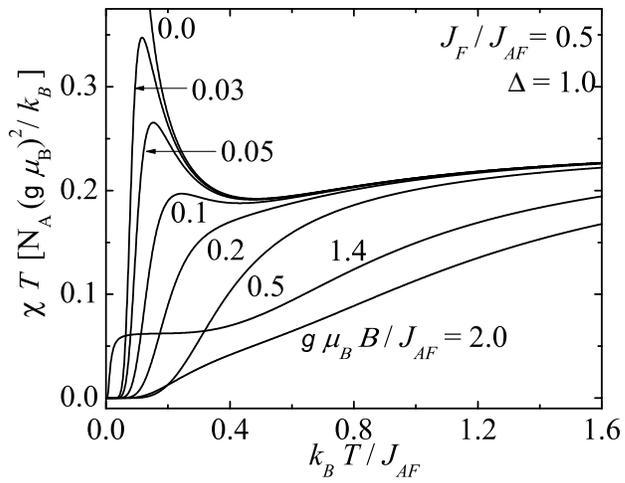
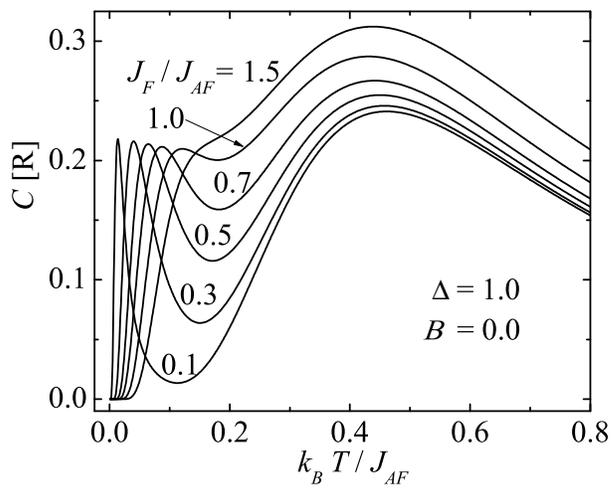
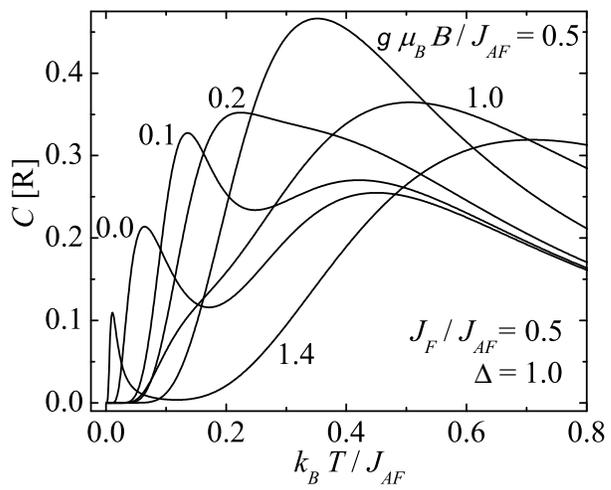
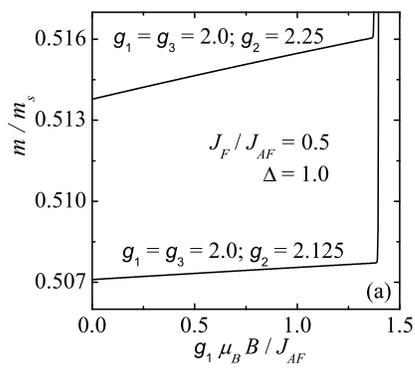
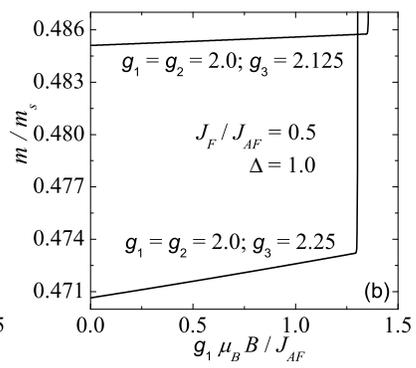
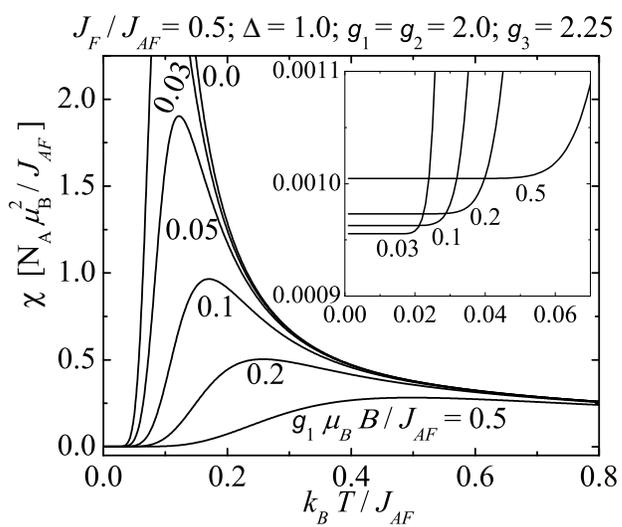
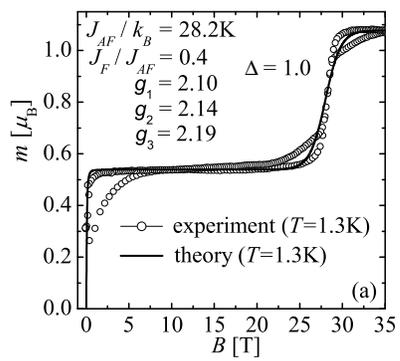
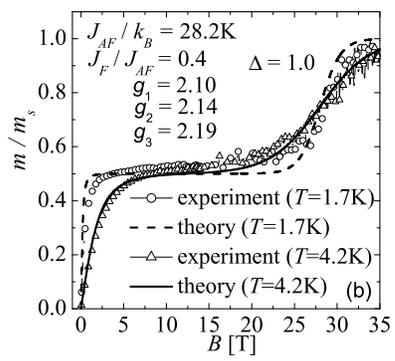
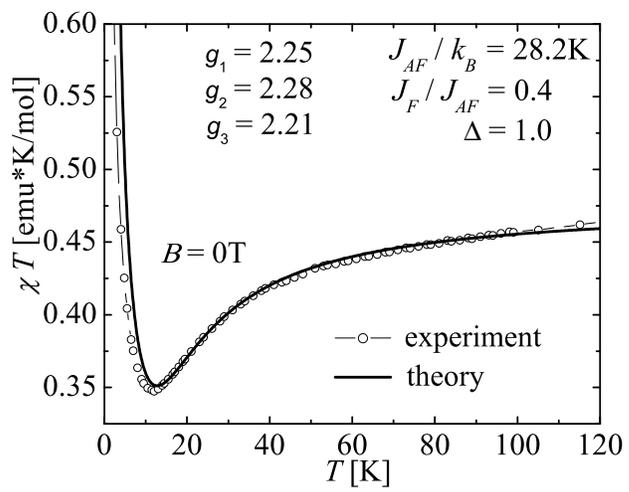
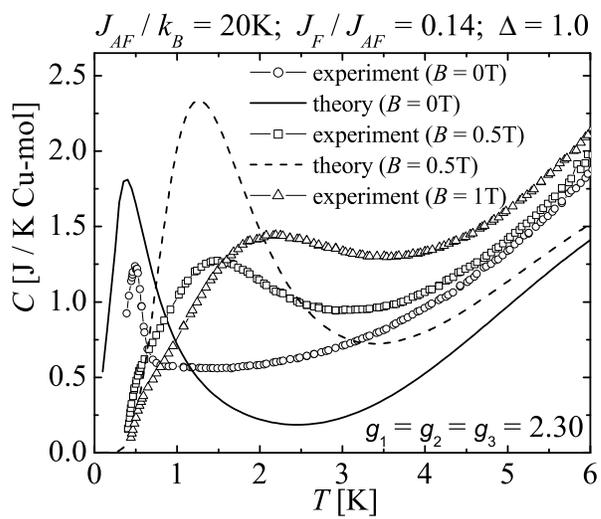
\end{document}